\documentclass[]{spie}  %>>> use for US letter paper
\usepackage[dvips]{graphicx}

\newcommand{\NH}{\mbox {$N_{\rm H}$}}

\newcommand{\name}{\mbox{1E~0102.2-7219}}

\title{The SMC SNR 1E0102.2-7219 as a Calibration Standard for X-ray
Astronomy in the 0.3-2.5 keV Bandpass  }

\author{Paul P. Plucinsky\supit{a}, 
Frank Haberl\supit{b},  
Daniel Dewey\supit{c}, 
Andrew P. Beardmore\supit{d}, 
Joseph M. DePasquale\supit{a},  
Olivier Godet\supit{d}, 
Victoria Grinberg\supit{b},
Eric D. Miller\supit{c},  
A.M.T. Pollock\supit{e},  
Steve Sembay\supit{d},  
Randall K. Smith\supit{g}  
\skiplinehalf
\supit{a}Harvard-Smithsonian Center for Astrophysics, 60 Garden St., 
Cambridge, MA, USA 02138 \\ 
\supit{b} Max-Planck-Institut f\"ur Extraterrestrische Physik, Giessenbachstra{\ss}e , 85748 Garching, Germany \\
\supit{c} MIT Kavli Institute for Astrophysics and Space Research, Cambridge, MA 02139 \\
\supit{d} Department of Physics and Astronomy, University of
Leicester, Leicester LE1 7RH \\
\supit{e} European Space Astronomy Centre, Apartado 78, Villanueva del Canada, 
28691 Madrid, Spain \\
\supit{g}Code 662 NASA Goddard Space Flight Center, Greenbelt, MD, 
USA 20771 \\ }

\authorinfo{Further author information: (Send correspondence to
Paul Plucinsky)\\Paul Plucinsky.: E-mail: plucinsky@cfa.harvard.edu \\ }
 
\begin{document} 
  \maketitle 

%%%%%%%%%%%%%%%%%%%%%%%%%%%%%%%%%%%%%%%%%%%%%%%%%%%%%%%%%%%%% 
\begin{abstract}

The flight calibration of the spectral response of CCD instruments 
below 1.5 keV is difficult in general because of the  lack of strong 
lines in the on-board calibration sources typically available.
We have been using \name\, the brightest supernova remnant in 
the Small Magellanic Cloud, to evaluate the response models of
the ACIS CCDs on the Chandra X-ray Observatory (CXO), the EPIC CCDs
on the XMM-Newton Observatory, the XIS CCDs on the {\em Suzaku} Observatory,
and the XRT CCD on the {\em Swift} Observatory.  E0102 has strong lines of 
O, Ne, and Mg below 1.5 keV and little or no Fe emission to 
complicate the spectrum.  
%The spectrum of E0102 has been
%well-characterized using the RGS grating instrument on XMM-Newton and
%the HETG grating instrument on Chandra.  We have used the high-resolution
%spectral data from both gratings instruments to develop a spectral
%model that is consistent with the data from both gratings instruments
%and which can be used to fit the CCD spectra.  
The spectrum of E0102 has been well characterized using
high-resolution grating instruments, namely the XMM-Newton RGS and the 
CXO HETG, through which a consistent spectral model has been developed
that can then be used to fit the lower-resolution CCD spectra.
Fits with this model
are sensitive to any problems with the gain calibration and the
spectral redistribution model of the CCD instruments. We have also
used the measured intensities of the lines to investigate
the consistency of the effective area models for the various
instruments around the bright O ($\sim570$~eV and $\sim654$~eV) and Ne 
($\sim910$~eV and $\sim1022$~eV) lines.
We find that the measured fluxes of the O~{\small VII}~triplet,
the O~{\small VIII}~Ly~$\alpha$ line, the Ne~{\small IX}~triplet, and
the Ne~{\small X}~Ly~$\alpha$ line generally agree to within $\pm10\%$
for all instruments, with 28 of our 32 fitted normalizations within
$\pm10\%$ of the RGS-determined value.  The maximum discrepancies,
computed as the percentage difference between the lowest and highest 
normalization for any instrument pair, are 23\% for 
the O~{\small VII}~triplet,  24\% for the O~{\small VIII}~Ly~$\alpha$ line,
13\% for the Ne~{\small IX}~triplet, and 19\% for the 
Ne~{\small X}~Ly~$\alpha$ line. If only the CXO and XMM are
compared, the maximum discrepancies are 22\% for 
the O~{\small VII}~triplet,  16\% for the O~{\small VIII}~Ly~$\alpha$ line,
4\% for the Ne~{\small IX}~triplet, and 12\% for the 
Ne~{\small X}~Ly~$\alpha$ line.

%This work was supported by NASA contract NAS8-03060.

\end{abstract}

%>>>> Include a list of keywords after the abstract 

\keywords{Chandra X-ray Observatory, XMM-Newton, Suzaku, Swift, ACIS,
EPIC, RGS, HETG, XIS, XRT,   
charge-coupled devices (CCDs), X-ray detectors, X-ray spectroscopy }

%%%%%%%%%%%%%%%%%%%%%%%%%%%%%%%%%%%%%%%%%%%%%%%%%%%%
\section{INTRODUCTION}
\label{sect:intro}  % \label{} allows reference to this section

  This paper reports the progress of a working group within the {\em 
International Astronomical Consortium for High Energy Calibration}
(IACHEC) to develop a calibration standard for X-ray astronomy in the 
bandpass from 0.3 to 2.5 keV.  A brief introduction to the IACHEC
organization, its objectives and meetings, may be found at the web
page {\tt http://www.iachec.org/}.  Our working group was tasked with
selecting celestial sources with line-rich spectra in the 0.3-2.5~keV
bandpass which would be suitable cross-calibration targets for the
current generation of X-ray observatories.  The desire for strong
lines in this bandpass stems from the fact that the
quantum efficiency and spectral resolution of the current CCD-based
instruments is changing rapidly from 0.3 to 1.5~keV and also 
significantly around the Si~K~edge, but the on-board
calibration sources currently in use typically have strong lines at
only two energies, 1.5~keV (Al~K$\alpha$) and 5.9~keV (Mn~K$\alpha$). 
The only option
available to the current generation of flight instruments to calibrate
any time variable response is to use celestial sources.  The missions which
have been represented in this work are the {\em Chandra X-ray
Observatory}\cite{weiss00,weiss02} (CXO), the {\em X-ray Multimirror
Mission}\cite{jansen2001} (XMM-Newton), the {\em ASTRO-E2 Observatory} ({\em
Suzaku}), and  the {\em Swift} Gamma-ray Burst Mission.  Data from the
following instruments have
been included in this analysis: the {\em High-Energy Transmission Grating}
(HETG) and the {\em Advanced CCD Imaging Spectrometer}\cite{bautz98,garmire03,garmire92} (ACIS) on the
CXO, the  {\em Reflection Gratings Spectrometers}\cite{denherder2001}
(RGS), the {\em European Photon Imaging Camera} (EPIC) 
{\em Metal-Oxide Semiconductor}\cite{turner2001} (MOS) CCDs and the 
EPIC p-n junction\cite{strueder2001} (pn) CCDs on XMM-Newton, the 
{\em X-ray Imaging Spectrometer} (XIS) on {\em Suzaku}, and the 
{\em X-ray Telescope}\cite{burrows2005,godet2007} (XRT) on {\em Swift}.

  Suitable calibration targets would need to possess the following 
qualities.  The source would need to be constant in time, 
to have a simple spectrum defined by a few bright lines
with a minimum of line-blending, and to be extended
so that ``pileup'' effects in the CCDs are minimized but not so
extended that the off-axis response of the telescope dominates
the uncertainties in the response.  Our working group focused on supernova
remnants (SNRs) with thermal spectra and without a central source such as a
pulsar, as the class of source which had the greatest likelihood of
satisfying these criteria.  We narrowed our list to the Galactic SNR
Cas-A, the Large Magellanic Cloud remnant N132D and the Small
Magellanic Cloud remnant \name\/.  We discarded Cas-A since it is a
relatively young ($\sim350$~yr) SNR with significant brightness
fluctuations in the X-ray, radio, and optical over the past three
decades, it contains a faint (but apparently constant) central source,
and it is relatively large (radius $\sim3.5$~arcminutes).  We
discarded N132D because it has a complicated, irregular morphology in
the X-rays and its spectrum shows strong, complex Fe emission.  We
therefore settled on \name\ as the
most suitable source given its relatively uniform morphology, small
size (radius $\sim0.4$~arcminutes), and comparatively simple X-ray
spectrum.

%-------------
   \begin{figure}[h]
%   \vspace{-0.50in}
  \begin{center}
  \begin{tabular}{c}
  \includegraphics[width=3.0in,angle=0]{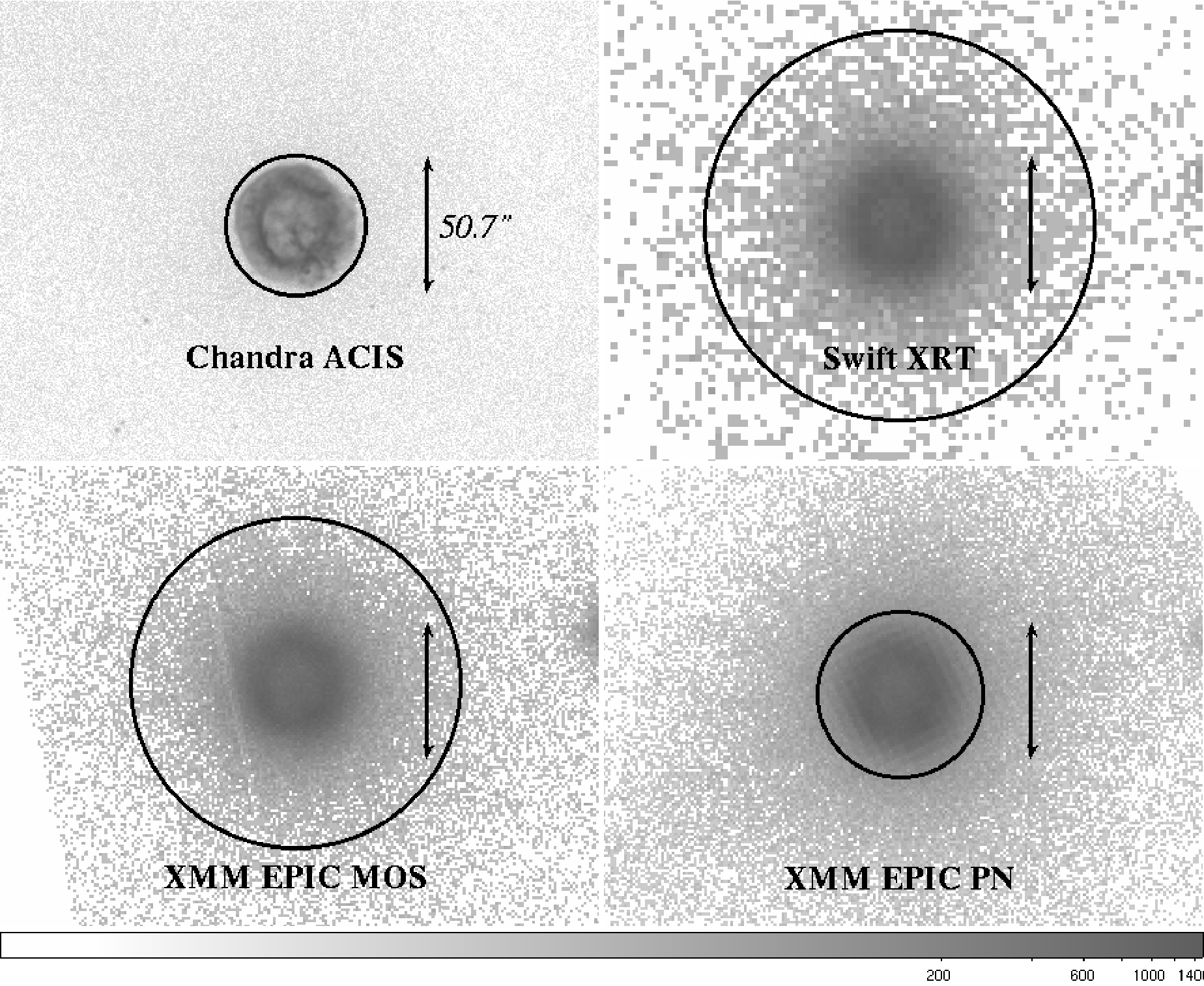}
  \includegraphics[width=3.0in,angle=0]{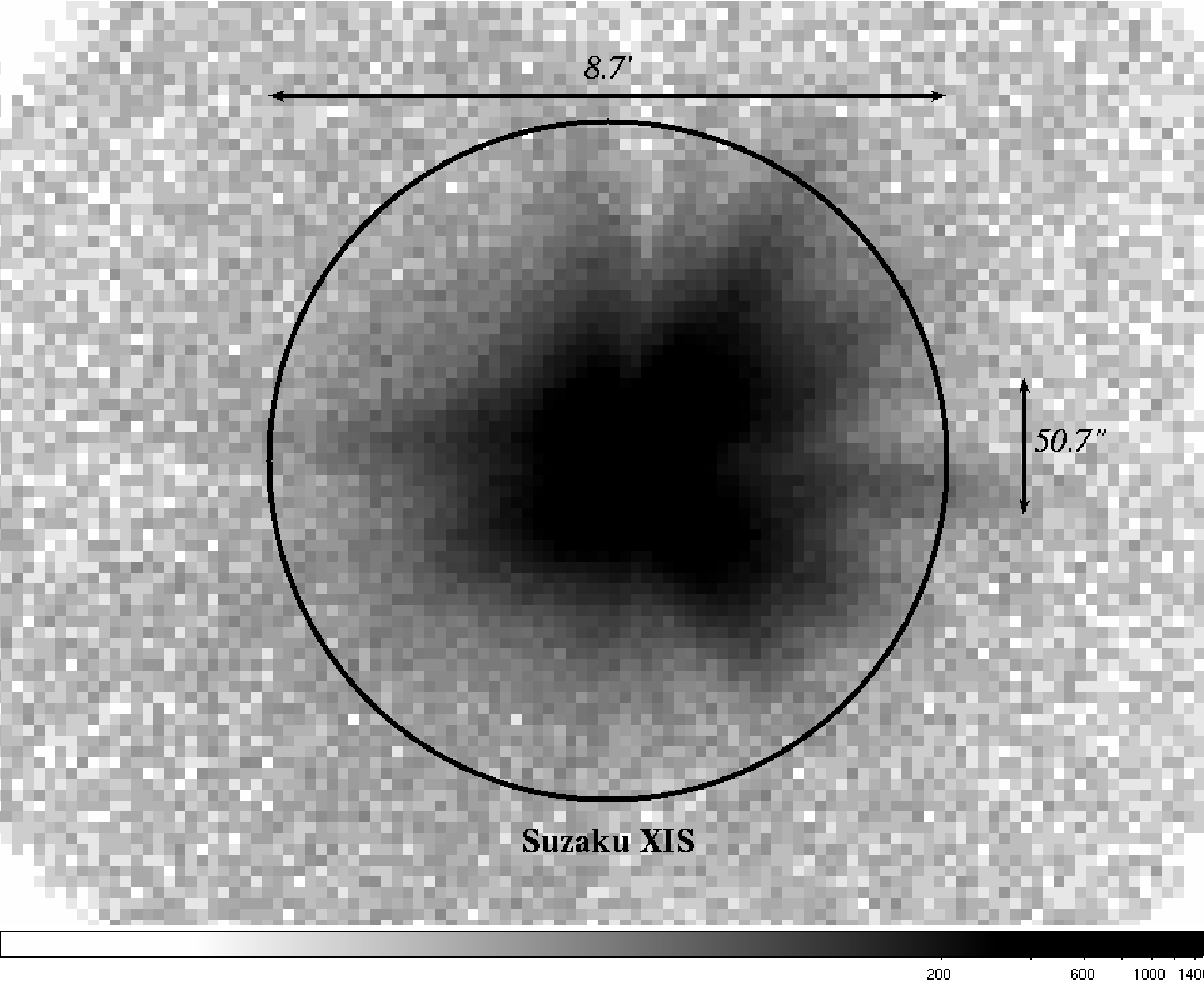}
  \end{tabular}
  \end{center}
  \vspace{-0.15in}
  \caption[image] 
%>>>> use \label inside caption to get Fig. number with \ref{}
  { \label{fig:image} Images of E0102 from ACIS S3 (top left),
MOS(bottom left), XRT(middle top), pn(middle bottom), and XIS(right). 
The black circles indicate the extraction regions used for the
spectral analysis. The high angular resolution of the CXO's
mirrors are evident in the structure resolved in the SNR and the small
extraction region.} 
  \end{figure} 
%-------------

\begin{figure}[h]
%   \vspace{-0.50in}
  \begin{center}
  \begin{tabular}{c}
  \includegraphics[width=4.9in,angle=270]{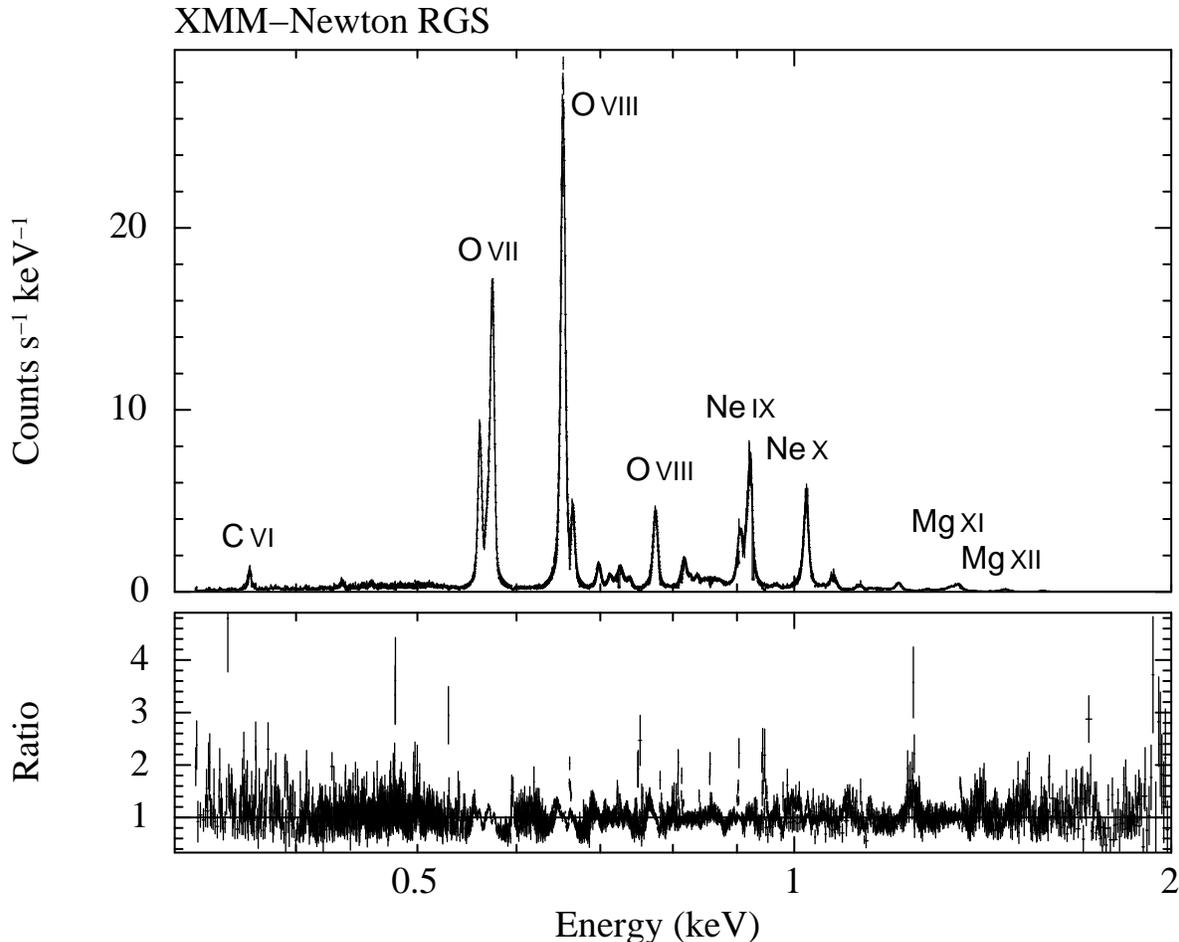}
  \end{tabular}
  \end{center}
  \vspace{-0.15in}
  \caption[image] 
%>>>> use \label inside caption to get Fig. number with \ref{}
  { \label{fig:rgs_spec_lin}  RGS1/2 spectrum of E0102 from a combination
of 23 observations. Note the bright lines of O, Ne, and Mg.
%See the electronic edition for a color version of this figure. 
}
  \end{figure}

  The SNR \name\ (hereafter E0102) was discovered 
by the {\em Einstein Observatory}\cite{seward1981}.  It is the
brightest SNR in X-rays in the {\em Small Magellanic Cloud} (SMC).
E0102 has been extensively imaged by CXO\cite{gaetz2000,hughes2000} and 
XMM-Newton\cite{sasaki2001}.  
Figure~\ref{fig:image} shows an image of E0102 with the relevant
spectral extraction region for each of the instruments included in this
analysis.  E0102 is classified as an
``O-rich'' SNR and has an estimated age of $\sim1,000$~yr.
The source diameter is small enough
such that a high resolution spectrum may be
acquired with the HETG on the CXO and the RGS on XMM-Newton.
The HETG
spectrum\cite{flanagan2004} and the RGS spectrum\cite{rasmussen2001}
both show strong lines of O, Ne, and Mg with little or no Fe.
E0102's spectrum is relatively simple compared to a typical SNR
spectrum.  
Figure~\ref{fig:rgs_spec_lin} displays the RGS spectrum from E0102.
The strong, well-separated lines in the energy range 0.5 to
1.5~keV make this source a useful target for calibration observations.
The source is extended enough to reduce the effects of photon pileup,
which distorts a spectrum. 
Although moderate pileup is expected in all the non-grating instruments 
when observed in modes with relatively long frame times.
The source is also bright enough to
provide a large number of counts in a relatively short observation.
% This image highlights the superb
%imaging capability of the optics on the CXO.

%-------------
%-------------

\section{CONSTRUCTION OF THE MODEL}
\label{sect:model} 

  Our objective was to develop a model which would be useful in
calibrating and comparing the response of the CCD instruments;
therefore, the model presented here is of limited value for
understanding E0102 as a SNR.  Our approach was to rely upon the
high-resolution spectral data from the RGS and HETG to identify and
characterize the bright lines and the continuum  
in the energy range from 0.3-2.0 keV and the moderate-spectral
resolution data from the MOS and pn to characterize the lines and
continuum above 2.0~keV.  Since our objective is
calibration, we decided against using any of the available plasma
emission models for several reasons. First, the CXO
results on E0102\cite{flanagan2004,gaetz2000,hughes2000} have shown
there are significant spectral variations with position in the SNR,
implying that the plasma conditions are varying throughout the
remnant.  Since the other missions considered here have poorer angular
resolution that the CXO, the emission from these regions 
is mixed such that the unambiguous interpretation of
the fitted parameters of a plasma emission model is difficult if not
impossible.  Second, the available parameter space in the more complex
codes is large, making it difficult to converge on a single best fit
which represents the spectrum.  We therefore decided to construct a
simple, empirical model based on interstellar absorption components, 
Gaussians for
the line emission, and continuum components which would be appropriate 
for our limited calibration objectives.

%-------------
   \begin{figure}[h]
%   \vspace{-0.50in}
  \begin{center}
  \begin{tabular}{c}
  \includegraphics[width=4.9in,angle=270]{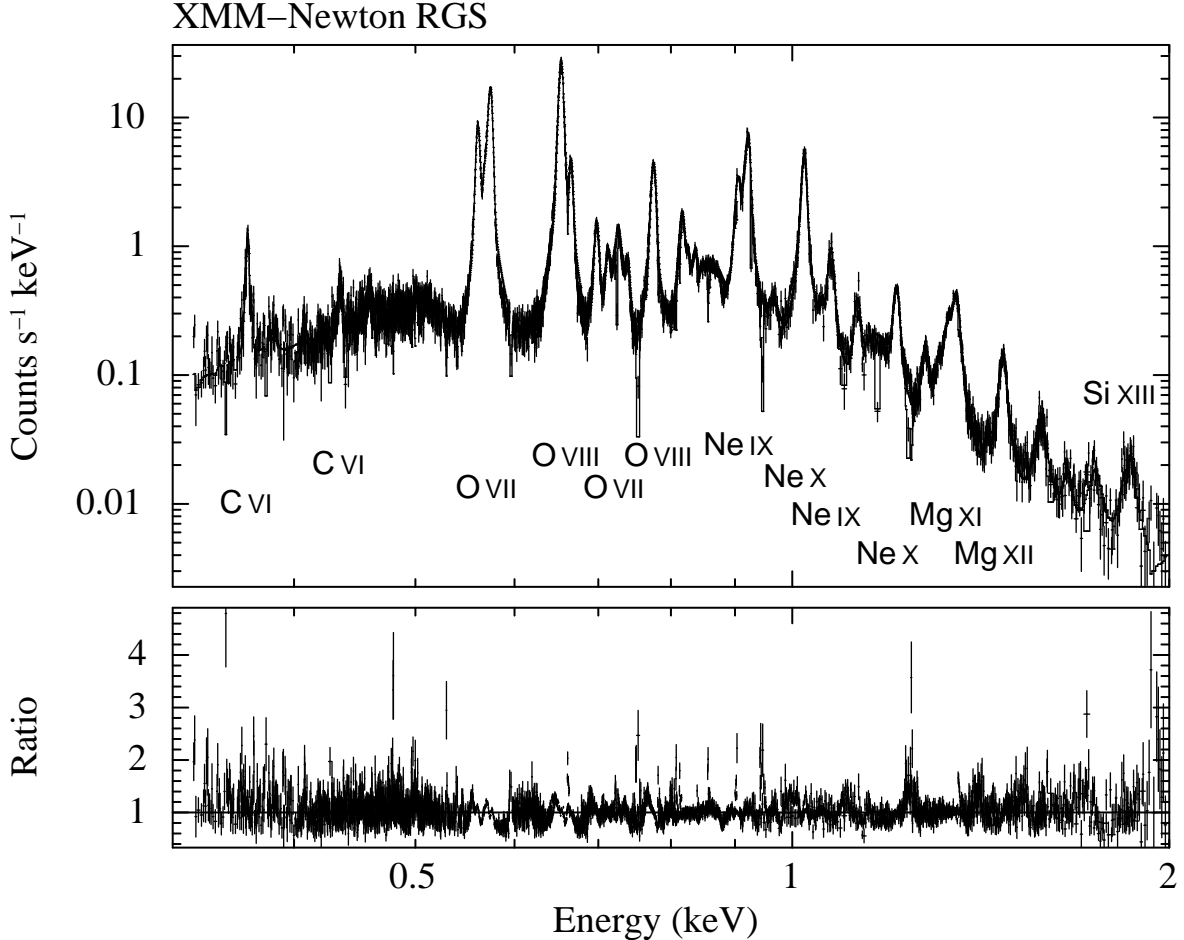}
  \end{tabular}
  \end{center}
  \vspace{-0.15in}
  \caption[image] 
%>>>> use \label inside caption to get Fig. number with \ref{}
  { \label{fig:rgs_spec_log} RGS1/RGS2 spectrum of E0102 from a combination
of 23 observations with a logarithmic Y axis to emphasize the
continuum and the weakest lines. 
%See the electronic edition for a color version of this figure.
}
  \end{figure} 
%-------------

   We assumed a two component absorption model using the 
{\tt tbabs}\cite{wilms2000} model in {\tt XSPEC}. The first component
was held fixed at $5.36\times10^{20}~{\rm cm^{-2}}$ to account for
absorption in the Galaxy.  The second component was allowed to vary
in total column, but with the abundances fixed to the lower abundances
of the SMC\cite{russell1989,russell1990,russell1992}.  
We modeled the continuum using a modified version of the {\tt APEC}\  
plasma emission model~\cite{smith2001} called the {\tt ``No-Line''}
model.   
This model excludes all line emission, while retaining all continuum  
processes including bremsstrahlung, radiative recombination continua  
(RRC), and the two-photon continuum from hydrogenic and helium-like  
ions (from the strictly forbidden ${}^2S_{1/2} 2s \rightarrow$\ gnd  
and ${}^1S_0 1s2s \rightarrow$\ gnd transitions, respectively).   
Although the bremsstrahlung continuum dominates the X-ray spectrum in  
most bands and at most temperatures, the RRCs can produce observable  
edges while the two-photon emission creates 'bumps' in specific energy  
ranges.  The {\tt ``No-Line''} model assumes collisional equilibrium 
and so may  
overestimate the RRC edges in an ionizing plasma or have the wrong  
total flux in some of the two-photon continua.  However, the available  
data did not justify the use of a more complex model, while the  
simpler bremsstrahlung-only model showed residuals in the RGS spectra  
that were strongly suggestive of RRC edges.
The RGS data were adequately fit by a single continuum component, but
the HETG, MOS, and pn data showed an excess at energies above 2.0~keV.
We therefore added a second continuum component to account for this
emission.

\begin{figure}[h]
%   \vspace{-0.50in}
  \begin{center}
  \begin{tabular}{c}
  \includegraphics[width=2.5in,angle=270]{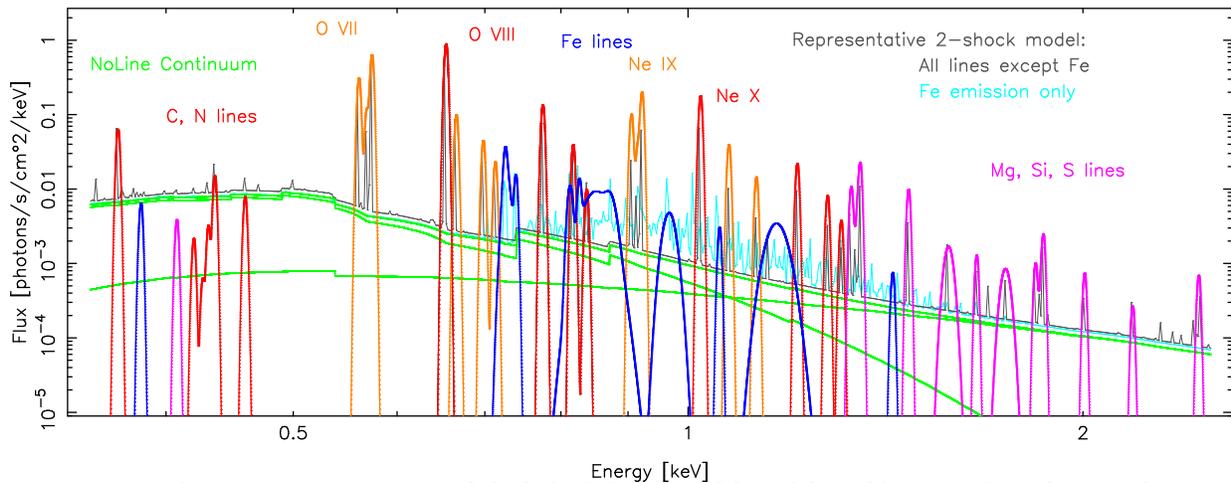}
  \end{tabular}
  \end{center}
  \vspace{-0.15in}
  \caption[HETG] 
%>>>> use \label inside caption to get Fig. number with \ref{}
  { \label{fig:vpshock} Line and continuum components of the hybrid
E0102 model used for calibration.  For reference, the emission predicted by a  
simple two-{\tt vpshock} model, scaled and summed with our continuum model,  
is shown. The calibration model clearly includes all expected bright lines as  
well as additional flux that may correspond to Fe emission from E0102.
For a color version of this figure see the on-line version. } 
  \end{figure}

The lines were modeled as simple Gaussians in {\tt XSPEC}.
The lines were identified in the RGS and HETG data in a hierarchical
manner, starting with the brightest lines and working down to
the fainter lines. 
We have used the ATOMDB~v1.3.1\cite{atomdb} database 
to identify the transitions which produce the observed lines. 
 The RGS spectrum from 23 observations totaling
708/680~ks for RGS1/RGS2 is shown in Figure~\ref{fig:rgs_spec_lin} 
with a linear Y axis
to emphasize the brightest lines.  The spectrum is dominated by the  
O~{\small VII}~triplet at 560-574~eV, the O~{\small VIII}~Ly~$\alpha$
line at 654~eV, the Ne~{\small IX}~triplet at 905-922~eV, and the
Ne~{\small X}~Ly~$\alpha$ at 1022~eV. This figure demonstrates  
the lack of strong Fe emission in the spectrum of E0102.  
  The
identification of the lines obviously becomes more difficult as the
lines become weaker.   Figure~\ref{fig:rgs_spec_log} shows the same
spectrum as in Figure~\ref{fig:rgs_spec_lin} but with a logarithmic
axis.  In this figure, one is able to see the weaker lines more
clearly and also the shape of the continuum.  Lines were added to the
spectrum at the known energies for the dominant elements, C, N, O, Ne,
Mg, Si, S, and Fe and the resulting decrease in the reduced $\chi^2$
value was evaluated to determine if the addition of the line was
significant.  The list of lines identified in the RGS and HETG data
were checked for consistency. 
The identified lines were 
compared against representative spectra from the {\tt vpshock} model
(with lines) to ensure that no strong lines were missed.
Figure~\ref{fig:vpshock} shows the comparison of our hybrid model of
continuum and lines to a representative spectrum from the {\tt
vpshock} model.

%-------------
%-------------

 In this manner a list of lines in the
0.3-2.0~keV bandpass was developed based upon the RGS and HETG data.
In addition, the temperature and normalization were determined for the
low-temperature {\tt APEC  ``No-Line''} continuum component.  These
model components were then frozen and the model compared to the pn,
MOS, and XIS data.  Weak lines above 2.0~keV were evident in the
pn, MOS,and XIS data and also what appeared to be an
additional continuum component above 2.0~keV.  Several lines were
added above 2.0~kev and a high-temperature continuum component with 
kT~$\sim1.7$~keV was added.
Once the model components above 2.0~keV had been determined, the RGS
data were re-fit with components above 2.0~keV frozen to these values
and the final values for the SMC \NH\/ and the low-temperature
continuum were determined. In practice, this was an iterative process
which required several iterations in fitting the RGS and MOS/pn/XIS
data.  Once the absorption and continuum components were determined,
the parameters for those components were frozen and the final
parameters for the line emission were determined from the RGS data.
We included 52 lines in the final model and these lines are described in
Table~\ref{tab:model}. When fitting the RGS data, the line energies
were allowed to vary by up to 1.0~eV from the expected energy to
account for the shifts when an extended source is observed by
the RGS.  Shifts of less than 1.0~eV are too small to be significant
when fitting the CCD instrument data.  The line widths were also
allowed to vary.  
%In most cases the line widths are small, less than
%$1.0\times10^{-2}$~keV, but in a few cases noted in the table the
%widths are larger than this value. 
In most cases the line widths are small but non-zero,
consistent with the Doppler widths seen in the RGS\cite{rasmussen2001}
and HETG\cite{flanagan2004} , $\sigma_E \approx 0.003 \times E$;  but
in a few cases noted in the table the widths are larger than this value.
This is mostly like due to weak,
nearby lines which our model has ignored. We do not have an
identification for the line-like feature at 1.4317~keV.

  As noted above, the identification of the lines becomes less certain
as the line fluxes get weaker.  Our primary purpose is to characterize
the flux in the bright lines of O, Ne, and Mg.  Any identification of
a line with flux less than $1.0\times10^{-4}~{\rm
photons~cm^{-2}~s^{-1}}$ in Table~\ref{tab:model} should be considered
tentative.  The Fe lines in  Table~\ref{tab:model} warrant special
discussion. There are nine Fe lines included in our model from
different ions.  We have not verified the self-consistency of the
Fe lines included in this model.  We anticipate that we will work on
this in the near future.  Of particular note is the Fe~{\small XIX}
line at 917~eV with zero flux.  We went through several iterations of
the model with this line included and excluded.  Unfortunately this
line is only 2~eV away from the Ne~{\small IX}~Intercombination line
at 915~eV and neither the RGS nor the HETG have the resolution to
characterize the emission in this region.  We have decided to
attribute all the flux in this region to the Ne~{\small
IX}~Intercombination line but have retained the Fe~{\small XIX} for
future investigations.  It is possible that some of the emission
which we have identified as Fe emission is due to other elements.  For
our calibration objective this is not important because all of the Fe
lines are weak and they do not have a significant effect on the fitted
parameters of the bright lines of O, Ne, and Mg.  We hope that future
instruments will have the resolution and sensitivity to uniquely
identify the weak lines in the E0102 spectrum.

\begin{table}[h]
\centering
%\large
\caption[]{{\bf {Spectral Lines Included in the E0102 Reference Model  
(v1.9)} }}

\label{tab:model}

\begin{tabular}{lllr|clllr}\hline\hline
Line ID & E (keV)\,$^{a}$ & $\lambda$~(\AA)\,$^{a}$ & Flux\,$^{b}$ &  &
Line ID & E (keV)\,$^{a}$ & $\lambda$~(\AA)\,$^{a}$ & Flux\,$^{b}$ \\
\hline
C~{\small VI~~Ly$\alpha$} & 0.3675 & {\small 33.737} & 175.2 & \quad &  
Ne~{\small IX~~i} & 0.9148 & {\small 13.553} & 249.6 \\
Fe~{\small XXIV} & 0.3826 & {\small 32.405} & 18.4 & \quad &  
Fe~{\small XIX} & 0.9172 & {\small 13.517} & 0.0 \\
S~{\small XIV} & 0.4075 & {\small 30.425} & 11.8 & \quad & Ne~{\small  
IX~~r} & 0.922 & {\small 13.447} & 1380.5 \\
N~{\small VI~~f} & 0.4198 & {\small 29.534} & 6.8 & \quad & Fe~{\small  
XX} & 0.9668\,$^{c}$ & {\small 12.824} & 120.5 \\
N~{\small VI~~i} & 0.4264 & {\small 29.076} & 2.0 & \quad & Ne~{\small  
X~~Ly$\alpha$} & 1.0217 & {\small 12.135} & 1378.3 \\
N~{\small VI~~r} & 0.4307 & {\small 28.786} & 10.5 & \quad &  
Fe~{\small XXIII} & 1.0564 & {\small 11.736} & 24.2 \\
C~{\small VI~~Ly$\beta$} & 0.4356 & {\small 28.462} & 49.5 & \quad &  
Ne~{\small IX~~He$\beta$} & 1.074 & {\small 11.544} & 320.7 \\
C~{\small VI~~Ly$\gamma$} & 0.4594 & {\small 26.988} & 27.3 & \quad &  
Ne~{\small IX~~He$\gamma$} & 1.127 & {\small 11.001} & 123.1 \\
O~{\small VII~~f} & 0.561 & {\small 22.1} & 1313.2 & \quad &  
Fe~{\small XXIV} & 1.168\,$^{c}$ & {\small 10.615} & 173.5 \\
O~{\small VII~~i} & 0.5686 & {\small 21.805} & 494.4 & \quad &  
Ne~{\small X~~Ly$\beta$} & 1.211 & {\small 10.238} & 202.2 \\
O~{\small VII~~r} & 0.5739 & {\small 21.603} & 2744.7 & \quad &  
Ne~{\small X~~Ly$\gamma$} & 1.277 & {\small 9.709} & 78.5 \\
O~{\small VIII~~Ly$\alpha$} & 0.6536 & {\small 18.969} & 4393.3 &  
\quad & Ne~{\small X~~Ly$\delta$} & 1.308 & {\small 9.478} & 37.1 \\
O~{\small VII~~He$\beta$} & 0.6656 & {\small 18.627} & 500.9 & \quad &  
Mg~{\small XI~~f} & 1.3311 & {\small 9.314} & 108.7 \\
O~{\small VII~~He$\gamma$} & 0.6978 & {\small 17.767} & 236.1 & \quad  
& Mg~{\small XI~~i} & 1.3431 & {\small 9.231} & 27.5 \\
O~{\small VII~~He$\delta$} & 0.7127 & {\small 17.396} & 124.9 & \quad  
& Mg~{\small XI~~r} & 1.3522 & {\small 9.169} & 231.0 \\
Fe~{\small XVII} & 0.7252 & {\small 17.096} & 130.9 & \quad & ~? 
~{\small ~?} & 1.4317 & {\small 8.659} & 8.1 \\
Fe~{\small XVII} & 0.7271\,$^{c}$ & {\small 17.051} & 165.9 & \quad &  
Mg~{\small XII~~Ly$\alpha$} & 1.4721 & {\small 8.422} & 110.2 \\
Fe~{\small XVII} & 0.7389 & {\small 16.779} & 82.3 & \quad &  
Mg~{\small XI~~He$\beta$} & 1.579\,$^{c}$ & {\small 7.852} & 50.6 \\
O~{\small VIII~~Ly$\beta$} & 0.7746 & {\small 16.006} & 788.6 & \quad  
& Mg~{\small XI~~He$\gamma$} & 1.659 & {\small 7.473} & 16.0 \\
Fe~{\small XVII} & 0.8124\,$^{c}$ & {\small 15.261} & 90.5 & \quad &  
Mg~{\small XII~~Ly$\beta$} & 1.745\,$^{c}$ & {\small 7.105} & 29.7 \\
O~{\small VIII~~Ly$\gamma$} & 0.817 & {\small 15.175} & 243.1 & \quad  
& Si~{\small XIII~~f} & 1.8395 & {\small 6.74} & 13.8 \\
Fe~{\small XVII} & 0.8258 & {\small 15.013} & 65.1 & \quad &  
Si~{\small XIII~~i} & 1.8538 & {\small 6.688} & 3.4 \\
O~{\small VIII~~Ly$\delta$} & 0.8365 & {\small 14.821} & 62.7 & \quad  
& Si~{\small XIII~~r} & 1.865 & {\small 6.647} & 34.6 \\
Fe~{\small XVIII} & 0.8503\,$^{c}$ & {\small 14.581} & 407.3 & \quad &  
Si~{\small XIV~~Ly$\alpha$} & 2.0052 & {\small 6.183} & 11.2 \\
Fe~{\small XVIII} & 0.8726\,$^{c}$ & {\small 14.208} & 89.6 & \quad &  
Si~{\small XIII~~He$\beta$} & 2.1818 & {\small 5.682} & 4.3 \\
Ne~{\small IX~~f} & 0.9051 & {\small 13.698} & 690.2 & \quad &  
S~{\small XV~~f,i,r} & 2.45 & {\small 5.06} & 12.7 \\
\hline

\end{tabular}

\begin{flushleft}
$^{a}$~Theoretical rest energies; wavelengths are $hc/E$. \\
$^{b}$~Observed flux in $10^{-6}$~photons~cm$^{-2}$\,s$^{-1}$ \\
$^{c}$~This line is broader than the nominal width, see text\\
\end{flushleft}

\end{table}

%-------------
   \begin{figure}[h]
%   \vspace{-0.50in}
  \begin{center}
  \begin{tabular}{c}
  \includegraphics[width=6.5in,angle=0]{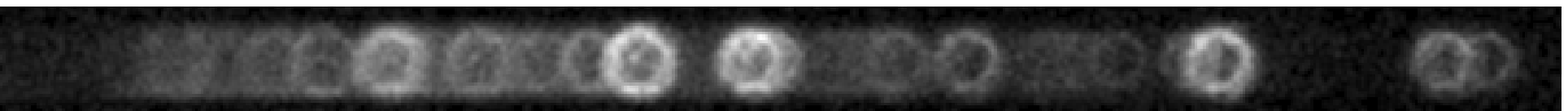} \\
  \includegraphics[width=6.5in,angle=0]{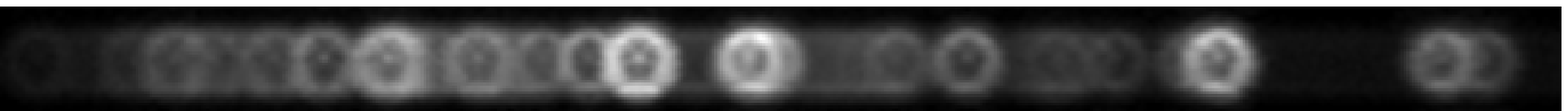}
  \end{tabular}
  \end{center}
  \vspace{-0.15in}
  \caption[MEG] 
%>>>> use \label inside caption to get Fig. number with \ref{}
  { \label{fig:MEG}    Images of E0102 MEG-dispersed data (top) and 
model (bottom).  The  four
MEG dispersed images ($\pm$1 orders from two epochs) have been combined
and displayed in the range from 4.6\AA\ to 23\AA (2.7--0.54 keV),  
clearly
showing bright rings of line emission for many lines.  The simulated
spectral image, below, is based on the nominal hybrid 
model.} 
  \end{figure} 
%-------------

   The number of free parameters needed to be significantly reduced
before fitting the CCD data in order to reduce the possible parameter
space.  In our fits, we have frozen the line energies and widths,
the SMC \NH\/, and the low-temperature {\tt APEC ``No-Line''}
continuum to the RGS-determined values.  The high-temperature
{\tt APEC ``No-Line''} component was frozen at the values determined from the
pn and MOS. The fixed absorption and continuum components are listed
in Table~\ref{tab:abs_cont}.
Since the CCD instruments lack the spectral resolution to 
resolve lines which are as close to each other as
the ones in the O~{\small VII}~triplet and the Ne~{\small IX}~triplet,
we treated nearby lines from the same ion as a ``line complex'' by
constraining the ratios of the line normalizations to be those
determined by the RGS and by constraining the line energies to the
known separations.  In practice, we would typically link the
normalization and energy of the Forbidden and Intercombination lines of
the triplet to the Resonance line (except for O~{\small VII} for which
we linked the other lines to the Forbidden line).  
Since we also usually freeze the
energies of the lines, this means that the three lines in the triplet
would have only one free parameter, the normalization of the Resonance
line.  We have constructed the model in {\tt XSPEC} such that it would
be easy to vary the energy of the Resonance line in the triplet (and
hence also the Forbidden and Intercombination) to examine the gain
calibration of a detector at these energies.  Our philosophy is to
treat 
nearby lines
as a  complex which can adjust together in normalization and energy.
In this paper, we focus on adjusting the normalization of the line
complexes only.  Since most of the power in the spectrum is
 in the bright line
complexes, we froze all the normalizations of the weaker lines.
The only normalizations which we allowed to vary were the 
O~{\small VII}~For, O~{\small VIII}~Ly~$\alpha$ line, the
Ne~{\small IX}~Res, and the Ne~{\small X}~Ly~$\alpha$ line
normalizations.  In addition, we found it useful to introduce a
constant scaling factor of the entire model to account for the fact
that the extraction regions for the various instruments were not
identical.  Finally, some of the instruments found it useful to allow
for a global gain shift.  In this manner, we restricted a model with
more than 200 parameters to have only 5 or 7 free parameters in our
fits.  The final version of this model in the {\tt XSPEC .xcm} file
format is available at the web page: 
{\tt
``http://cxc.harvard.edu/twiki/bin/view.cgi/SnrE0102/FinalModel''}.

\begin{table}[h]
\centering

\caption[ ]{\bf {Fixed Absorption and Continuum Components}}

\label{tab:abs_cont}
\begin{tabular}{|l|l|}
\hline
%\hline
%OBSID & Instrument & DATE & Exposure(s) & Mode/Notes \\
%\hline
%%\multicolumn{2}{l}{\bf BI(S3) Datasets } & & &    \\
%\hline
Galactic absorption  & \NH\/$=5.36\times10^{20}~{\rm cm^{-2}}$ \\
SMC absorption       & \NH\/$=5.76\times10^{20}~{\rm cm^{-2}}$ \\
{\tt APEC ``No-Line''} temperature \#1 & kT=0.164~keV \\
{\tt APEC ``No-Line''} normalization \#1 & $3.48\times10^{-2}~{\rm cm^{-5}}$ \\
{\tt APEC ``No-Line''} temperature \#2 & kT=1.736~keV \\
{\tt APEC ``No-Line''} normalization \#2 & $1.85\times10^{-3}~{\rm cm^{-5}}$ \\\hline

\end{tabular}

\end{table}

\section{OBSERVATIONS}
\label{sect:observations} 

 E0102 is routinely observed by the CXO, XMM-Newton and {\em Suzaku},
as a calibration target to monitor the response at energies below 1.5~keV.
For this paper, we have selected a subset of these observations for
each instrument since E0102 has been observed in different modes for
each instrument.  We have selected data from the instrument mode for
which we are the most confident in the calibration.  We now describe
the data processing and calibration issues for each instrument
individually.

{\bf RGS: } From all 23 XMM-Newton observations of E0102 between
16 April 2000 and 26 Oct 2007, we extracted spectra from the RGS 
instruments. For the complete analysis RGS1 and RGS2 were handled 
separately. The data were processed with XMM-SAS v7.1.0 using {\tt rgsproc} 
for the extraction of the spectra and the generation of the spectral 
response files. We used the task {\tt rgscombine} to add up the spectra and 
response files. This yields a total net exposure of 708080 s for RGS1 
and 680290 s for RGS2.

{\bf HETG: }  E0102 was observed with the HETG at two epochs: 
Sept.--Oct.\ 1999 and
in December of 2002; note that the first epoch is very early in the
CXO mission and at a focal plane temperature of $-110^{\circ}$\,C.
Analysis of the archive-retrieved data was carried out with
{\tt TGCat ISIS} 
scripts\footnote{ {\tt http://space.mit.edu/cxc/analysis/tgcat/} }
that called current {\tt CIAO} tools.
Of note in the analyses: the origin was set at the bright inner spot  
of the Q-stroke feature; zeroth-order radius was 60 pixels;
events were extracted
from widths of $\pm\,60$ pixels ($\pm\,0.0096$ in {\tt tg\_d}); and
order sorting values of 0.25 and 0.20 were used for the first
and second epochs, respectively.

The HETG view of E0102 is presented by the first epoch
observations \cite{flanagan2004}.
Because of the extended nature of E0102, non-standard analysis
methods are needed \cite{dewey2002} to get information from the
2D dispersed spectral images.  Here, ``Event2D'' 
software\footnote{ {\tt http://space.mit.edu/hydra/E2D\_demo/e2d\_demo.html} }
is used to take the E0102 hybrid spectral model, the {\tt CIAO}-generated  
{\tt arf},
and E0102 spectral images to create a Monte Carlo simulation of  
each
observation-grating-order data set.  Figure~\ref{fig:MEG} shows
the data and simulated model for the combined MEG data sets.
The second epoch data and model were compared using the number of counts
in six wavelength ranges covering the 0.54~keV to 1.58~keV high-counts
range of the MEG plus and minus first orders.
This crude binning provides for a quantitative comparison at the level
of the CCD analyses.  The resulting 12 ratios
of data/model and their errors were then used to determine the frozen $ 
\chi^2$ and
then the best fit values for the 5 free model parameters; these
results are given in the Tables~{\ref{tab:fit1} and~{\ref{tab:fit2}.

{\bf MOS:}  The EPIC MOS data were taken from two observations in XMM revolution 0065 and 
0247. Both MOS1 and MOS2 cameras were in Large Window mode (central CCD has 
$300\times300$ pixels with a frame time of 0.9s) with the thin filter in each 
observation. 
%The data were cleaned by excluding intervals of high background 
%caused by soft proton flaring. This procedure produced approximately 16.8 and 
%12.5 ksecs of cleaned data from each observation respectively. 
The source region was chosen to be a circle of radius 60" centered on the 
remnant. A background region of the same size was selected within the same 
central CCD. Response files were calculated using the XMM Science Analysis 
Software (SAS) task {\tt rmfgen} and {\tt arfgen}. As the source is
not a point source 
a map of the remnant in detector coordinates was used as an input
image to {\tt arfgen}
so that the calculated arf is weighted by the differential vignetting across 
the remnant.

In the brightest part of the remnant, the observed per-pixel count rate is 
approximately 0.005 ${\rm  cts~s^{-1}~pixel^{-1}}$. The 1\% pile-up
limit for the MOS is 0.0012~${\rm  cts~s^{-1}~pixel^{-1}}$ for an
extended source in Large Window mode so some 
moderate pile-up may be expected. We have applied a first order pile-up 
correction by using the information from the diagonal bi-pixel events in the 
MOS event files. The details of the pileup correction are documented
in the memo XMM-SOC-CAL-TN-0036 available
from the ESAC website at: \\
{\tt http://xmm2.esac.esa.int/external/xmm\_sw\_cal/calib/documentation/index.shtml\#EPIC}.

The data were then compared against the model for E0102.  It was found
that a significant improvement in the fit statistic was achieved in both 
cameras with a small gain shift. The fit statistics including the gain fit 
parameters are given in Table~\ref{tab:fit1}. 
%As per the agreed
%recipe, the uncer command was run after re-freezing the constant and 
%gain parameters and issuing a fit command.

{\bf pn:  } Observations of 1E0102 with EPIC pn were performed in full-frame, 
large window and small window mode using all available filters. To 
completely avoid pile-up effects we used only data from small window mode.
The data were processed with XMM-SAS v7.1.0 without utilizing the most
recent up date of the long-term CTI calibration adjustment available 
with EPN\_CTI\_0017.CCF which became public on 12 April 2008. 
Spectra were extracted using single-pixel events from a circular
region around the supernova remnant with a radius of 30
arcsec. Background spectra were extracted from a nearby circular
region with the same size. Response files were generated using 
{\tt rmfgen} and {\tt arfgen}, assuming a point source for PSF 
corrections.

{\bf ACIS:} The high angular resolution of the CXO compared to the
other observatories is apparent in Figure~\ref{fig:image}.
Unfortunately the bright parts of E0102 are significantly piled-up when
ACIS is operated in its full-frame mode with 3.2~s exposures. Also
unfortunately, the majority of the observations of E0102 over the
mission have been executed in full-frame mode.  There have been four subarray
observations of E0102, two in node 1 and two in node 0 of the
Backside-illuminated (BI) CCD S3.  We have
selected two OBSIDs from node 1 since the calibration in node 1 is
superior to that of node 0.  The data were processed with the CXO analysis
SW {\tt CIAO} v4.0 and the CXO calibration database CALDB~v3.4.3.
There are several time-dependent effects which the analysis SW
attempts to account for\cite{plucinsky2003,marshall2004,depasq2004}.  
The most important of these is the
efficiency correction for the contaminant on the ACIS optical-blocking
filter which significantly reduces the efficiency at energies around
the O lines.  The analysis SW also corrects for the CTI of the BI CCD
(S3), including the time-dependence of the gain.

{\bf XIS: } The XIS observations include 106 ksec of data from XIS1 (the BI
device) and 60 ksec of data from XIS0 (one of three FI devices).  This
represents the longest single observations of E0102 by {\em Suzaku}.  The
data were taken shortly after launch on 17 Dec 2005, when the
molecular OBF contamination was still relatively low, and when
spaced-row charge injection (SCI) was not yet being used.  Normal
observing mode was used, and 3x3 and 5x5 event editing mode were
combined in the dataset.

The data were processed with v2.0.6.13 of the XIS pipeline.
Spectra were extracted from a 4.35 arcmin (6 mm) aperture (the default
for a point source), and background spectra were extracted from a
surrounding annulus.  Response files were produced with the {\em Suzaku}
{\tt FTOOLS xisrmfgen} (v2007-05-14) and {\tt xissimarfgen}
(v2007-07-16).  The
ARF includes absorption due to OBF contamination, using the
contamination parameters as of 2007-12-24.

An additional point source lies 2 arcmin from E0102, well within the
spectral extraction region.  This X-ray binary (RXJ0103.6-7201) is
well-modeled by a power law with spectral index 0.9 plus a thermal
{\tt mekal} component with kT = 0.15 keV, with a strong correlation between
the component normalizations\cite{haberl2005}.  The emission
from this source dominates the extracted spectra above 3 keV; below 2
keV, E0102 thermal emission dominates by several orders of magnitude.
We have included the source in our spectral modeling, allowing a
single normalization to vary due to the observed source variability.

{\bf XRT:} The {\it Swift} XRT observations presented here were
restricted to Photon 
Counting mode data taken from 18~Feb~2005
to 22~May~2005 (observation ID 
numbers 00050050004,5,8,9), which were obtained before a micrometeoroid 
impact damaged two central columns on the CCD (on 27~May~2005) and before 
the effects of CTI became more noticeable from Oct~2005 onwards.

The data were processed using the latest release of the {\it Swift} 
software tools (version 2.8, release date 2007-12-07). After the standard 
data screening was applied the final exposure was 24.2ks. We selected 
grades 0-12 for the spectral comparison.
A circular region of radius 70.7 arcseconds was used for the spectral 
extraction. There is at present no XRT software tool capable 
of generating extended source ARF files, however the source was positioned 
near the centre of the CCD during the observations, so the effects of 
vignetting were negligible, and the extraction region was judged to 
contain 95 percent of the EEF. The RMF and ARF calibration files used in 
the spectral fitting were version 011, which will be available in the 
next calibration release.

%  Table~\ref{tab:obs} lists the relevant information for these
%observations.  There are many more observations of
%E0102 with the CXO and XMM-Newton, however, they were conducted in
%modes in which pileup is significant and/or the instrument configuration
%is significantly different to prevent a straightforward comparison
%(ie: the use of different filters for the XMM-Newton instruments).

\begin{table}[h]
\centering

\caption[ ]{\bf {Observations Used for this Analysis}}

\label{tab:obs}
\begin{tabular}{|c|l|l|r|r|l|}
%\hline
\hline
OBSID & Instrument & DATE & Exposure(s) & Counts & Mode
\\
 &  &  &  & (0.5-2.0 keV) &  \\
\hline
%%\multicolumn{2}{l}{\bf BI(S3) Datasets } & & &    \\
%\hline
3828    & ACIS-HETG & 2002-12-20 & 137660.0 & 49600 & TE, Faint, 3.2 s frametime
\\
3545    & ACIS-S3 & 2003-08-08 & 7862.1 &  55160 & TE, 1/4 subarray, 1.1~s frametime \\
6765    & ACIS-S3 & 2006-03-19 & 7635.5 & 50067 & TE, 1/4 subarray, 0.8~s frametime \\
0412980301 & pn   & 2007-10-26 & 25690.7 & 275397 & sm win, 6~ms
frametime, med filter \\  
0123110201 & MOS1  & 2000-04-16      & 16784.5  & 58007  & large win, 0.9~s
frametime, thin filter \\
0123110201 & MOS2  & 2000-04-16      & 16786.7  & 56235 & large win, 0.9~s
frametime, thin filter \\
0135720601 & MOS1  & 2001-04-14     & 12547.0  & 43626 & large win, 0.9~s
frametime, thin filter \\
0135720601 & MOS2  & 2001-04-14     & 12548.9  & 41575 & large win, 0.9~s
frametime, thin filter \\
100044010 & XIS0  & 2005-12-17 & 60000.0  & 96948 & normal,3x3+5x5,
SCI off   \\
100044010 & XIS1  & 2005-12-17 & 106000.0 & 342505 & normal,3x3+5x5,
SCI off   \\
00050050004 & XRT & 2005-02-18 &  24200  & 26254 & combination of 4
observations \\
\hline

\end{tabular}

\end{table}

\section{RESULTS}
\label{sect:results} 

The spectra for each of the CCD instruments were first compared to the
RGS-determined model without allowing any of the parameters to vary.
The resulting values of the $\chi^2$ are listed in
Table~\ref{tab:fit1} in the columns labelled ``Norms Frozen''.
It was noticed during this comparison that the MOS, ACIS, and XIS
spectra showed similar residuals in that the model appeared higher
than the data at low energies around the O lines.  It was also noticed
that a global offset to account for different size extraction regions
would help to reconcile the overall normalization of the spectra.
The spectra were then fit with 5 free parameters (with the exception
of the MOS and XIS fits in which two more parameters were allowed to
vary to account for a global gain offset).  
The five free parameters are the constant factor to
multiply each spectrum by and the O~{\small VII} triplet, 
O~{\small VIII}~Ly~$\alpha$, Ne~{\small IX}~triplet \& 
Ne~{\small X}~Ly~$\alpha$ line normalizations.  The fits were done first
with the constant factor free.  The constant factor was then frozen
and the spectra refit and the 90\% confidence limit (CL) on the line 
normalizations
determined.  Table~\ref{tab:fit1} lists the $\chi^2$ values
for these fits in the columns labelled ``Norms Free''.  Note that the
HETG data were fit in six broad energy bands unlike the CCD
instruments.  None of the
fits are formally acceptable and the quality of the fit around the
bright lines varies.
The spectral fits with the line normalizations free are
displayed in the following figures: 
Figure~\ref{fig:acis_spec} shows the ACIS data, 
Figure~\ref{fig:mos1_spec} shows the MOS data, 
Figure~\ref{fig:pn_spec} shows the pn data, 
Figure~\ref{fig:xis_spec} shows the XIS data, 
Figure~\ref{fig:xrt_spec} shows the XRT data.
The fit to the ACIS data has a reduced $\chi^2$ of 1.72 and appears to
fit the data reasonably well with the largest residuals below 0.5~keV
and above 1.5~keV.  Therefore, the bright O and Ne lines appear to be
well-fitted.  The MOS1 data have the lowest reduced $\chi^2$ of 1.57
and the data appear to be well-fitted with some small systematics in
the residuals. The MOS2 data are not fit as well with a reduced 
$\chi^2$ of 1.96.  The pn data are fitted with a reduced $\chi^2$ of
3.32 with large residuals on the low-energy side of the  O~{\small
VII}~triplet. Nevertheless, the O~{\small VIII} Ly~$\alpha$ line, 
the Ne~{\small IX} Res line, and the Ne~{\small X}~Ly~$\alpha$ line 
appear to be well-fitted.  The XIS1 spectral fit has the worst 
reduced $\chi^2$ of
6.40 with large residuals below 0.5~keV and above 1.5~keV.  But the
residuals around the O and Ne lines are relatively small.  Finally,
the XRT data are fit with a reduced $\chi^2$ of 2.32, with large
residuals above 1.5~keV presumably due to pileup.

\begin{table}[h]
\centering

\caption[ ]{\bf {Fit Statistics for the RGS model before and after the
Normalizations are Allowed to Vary}}

\label{tab:fit1}
\begin{tabular}{|l|c|c|c|c|c|c|}
%\hline
\hline
 & \multicolumn{2}{l}{\bf Norms Frozen } & \multicolumn{2}{l}{\bf
Norms Free } &   &  \\
Instrument & DOF  & $\chi^2$ & DOF  & $\chi^2$  & Gain  Offset(keV) &
Gain Slope\\
\hline
%%\multicolumn{2}{l}{\bf BI(S3) Datasets } & & &    \\
%\hline
HETG & 12 & 42.7 & 7 & 14.1 &  &  \\
\hline
ACIS & 253 & 714.1 & 249 & 427.3 &  &  \\
\hline
MOS1 & 318 & 719.1 & 315 & 492.9  & $-1.41\times10^{-3}$  & 1.005  \\        
\hline
MOS2 & 315 & 933.1 & 312  & 611.7  & $-1.38\times10^{-3}$  & 1.006   \\        
\hline
pn   & 365 & 1292.5 & 362 & 1202.2 & & \\
\hline
XIS0   & 228  & 884.7  & 228 & 489.2 & $-8.82\times10^{-03}$ & 1.006 \\
\hline
XIS1   & 228  & 2747.1 & 228 & 1456.9 & $-7.62\times10^{-03}$ & 1.009
\\
\hline
XRT  & 119 & 352.1  & 116 & 269.1 & & \\
     
\hline

\end{tabular}

\end{table}

%-------------
   \begin{figure}[h]
%   \vspace{-0.50in}
  \begin{center}
  \begin{tabular}{c}
  \includegraphics[width=4.0in,angle=270]{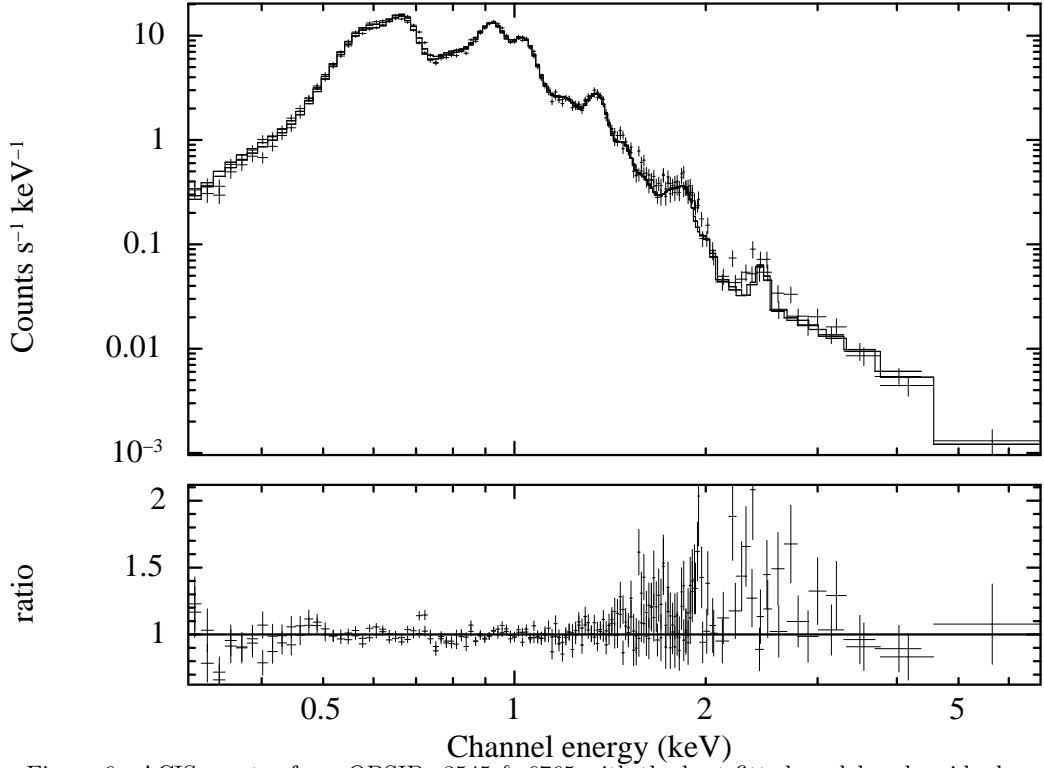}
  \end{tabular}
  \end{center}
  \vspace{-0.15in}
  \caption[ACIS] 
%>>>> use \label inside caption to get Fig. number with \ref{}
  { \label{fig:acis_spec} ACIS spectra from OBSIDs 3545 \& 6765 with
the best-fitted model and residuals.} 
  \end{figure} 
%-------------

\begin{table}[h]
\centering

\caption[ ]{\bf {Fitted Values for Constant Factor and Line Complex Normalizations }}

\label{tab:fit2}
\begin{tabular}{|l|c|c|c|c|c|}
%\hline
\hline
Instrument & Constant & O~{\small VII} For & O~{\small
VIII}~Ly~$\alpha$ & Ne~{\small IX} Res & Ne~{\small X}~Ly~$\alpha$  \\
           &          & Norm  & Norm & Norm & Norm \\
           &          & ($10^{-3} {\rm ph~cm^{-2} s^{-1}}$) 
& ($10^{-3} {\rm ph~cm^{-2} s^{-1}}$) 
& ($10^{-3} {\rm ph~cm^{-2} s^{-1}}$)
& ($10^{-3} {\rm ph~cm^{-2} s^{-1}}$) \\
\hline
%%\multicolumn{2}{l}{\bf BI(S3) Datasets } & & &    \\
%\hline
RGS1     & 1.00 & 1.313   &      4.393         & 1.381        & 1.378 \\
\hline
RGS2     & 0.96 & 1.313   &      4.393         & 1.381        & 1.378 \\
\hline
HETG    & 1.012 & 1.370   &      4.567         & 1.323         & 1.375  \\
90\%CL &   & [1.320,1.419] & [4.457,4.676] & [1.299,1.347] & [1.354,1.397] \\
Scaled  &       & 1.386   &      4.622         & 1.339       & 1.392 \\
\hline
ACIS    & 1.010 & 1.130   &      4.098         & 1.337        & 1.400 \\
90\%CL &   & [1.107,1.154] & [4.038,4.158] & [1.316,1.358] & [1.374,1.425] \\
Scaled  &       & 1.141   &      4.139     &     1.350       &  1.414  \\  
\hline
MOS1    & 0.993 & 1.296   &      4.384         & 1.354       &  1.407       \\
90\%CL &   & [1.271,1.320] & [4.323,4.446] & [1.335,1.373] & [1.384,1.429]  \\
Scaled  &      &  1.287   &      4.353     &     1.345       &  1.397  \\  
\hline
MOS2    & 1.016 & 1.305   &      4.318         & 1.333       &  1.356       \\
90\%CL &   & [1.280,1.323] & [4.256,4.379] & [1.314,1.353] & [1.333,1.378]  \\
Scaled  &       & 1.326      &   4.387       &   1.354     &    1.378 \\  
\hline
pn      & 0.942 & 1.366   &      4.235         & 1.388       &  1.340       \\
%90\%CL  &    & [$\pm0.01$] & [$\pm0.035$] & [$\pm0.014$] & [$\pm0.017$]  \\
90\%CL &  & [1.354,1.378] & [4.200,4.270] & [1.374,1.402] &
[1.323,1.356] \\
Scaled  &    &    1.287   &      3.989         & 1.307      &  1.262   \\  
\hline
XIS0    & 1.013 & 1.231   &  4.047        & 1.333   &   1.368       \\
90\%CL &  & [1.187,1.274] & [3.971,4.122] & [1.315,1.351] & [1.348,1.387] \\
Scaled  &    &    1.247  &   4.100        & 1.350   &   1.386   \\  
\hline
XIS1    & 1.024 &  1.142       &  4.006        &  1.418        &   1.463  \\
90\%CL &  & [1.130,1.155] & [3.975,4.038] & [1.407,1.429] & [1.450,1.476] \\
Scaled  &    &     1.169   &      4.102        &  1.452         &  1.498   \\  
\hline
XRT    & 0.934  &  1.206       &  4.001        & 1.375      & 1.551  \\
90\%CL & & [1.138,1.262] & [3.873,4.198] & [1.340,1.446] & [1.511,1.631] \\
Scaled   &    &     1.126   &      3.737        &   1.285       &   1.448 \\

\hline

\end{tabular}

\end{table}

%-------------
   \begin{figure}[h]
%   \vspace{-0.50in}
  \begin{center}
  \begin{tabular}{c}
  \includegraphics[width=1.7in,angle=270]{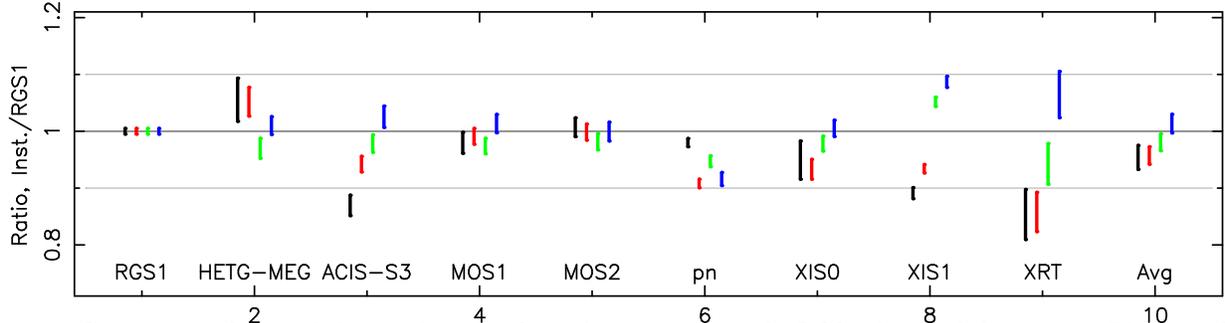}
  \end{tabular}
  \end{center}
  \vspace{-0.15in}
  \caption[pn] 
%>>>> use \label inside caption to get Fig. number with \ref{}
  { \label{fig:comp_norms} Comparison of the scaled normalizations for
each instrument to the RGS values and the average.  There are four
points for each
instrument which are from left to right, O~{\small VII}, O~{\small
VIII}, Ne~{\small IX}, \& Ne~{\small X}. The length of the line
indicates the 90\% CL for the scaled normalization.
 See the on-line version for
a color figure.
}
  \end{figure} 
%-------------

 The fitted line normalizations for the O~{\small VII}~For line, the
O~{\small VIII}~Ly~$\alpha$ line, the Ne~{\small IX}~Res line, \& 
Ne~{\small X}~Ly~$\alpha$ line are listed in Table~\ref{tab:fit2}.
The first and second rows include the results for the RGS1 and RGS2.
Note that the line normalizations are the same but the scale factor is
0.96 for RGS2, indicating that there is a systematic 4\% offset
between RGS1 and RGS2 . The results for the other instruments are then
presented in groups of three rows.  Within a group of three rows for a
given instrument,
the first row gives the best-fitted value, the second row gives the 
90\% CL, and the third row gives the ``scaled'' value where the
best-fitted value has been multiplied by the constant factor.
Figure~\ref{fig:comp_norms} presents the data in Table~\ref{tab:fit2}
in a graphical manner. The scaled normalizations for each instrument
are compared to the RGS-determined values and the average of all
instruments is also plotted.  Several patterns are clear in this plot.
First, ACIS, XIS, and XRT show the same trend of increasing
normalization with energy.  The MOS1 and MOS2 agree well with the
RGS1, but it should be noted that the MOS quantum efficiency was
adjusted in 2007 with the goal of improving the agreement with RGS.
If the MOS data are analyzed with the previous quantum efficiency
model, the pattern in the normalizations looks similar to ACIS, XIS,
and XRT. Finally, the pn data appear to be low relative to the RGS1 by
about 5\%.  In general, the scaled normalizations for all instruments
agree with the RGS values with $\pm10\%$.  Specifically, 28 of the 32
normalizations are within  $\pm10\%$ of the RGS1 values.
The scaled normalizations agree to
within 14\% and 19\% for the  Ne~{\small IX}~Res and Ne~{\small
X}~Ly~$\alpha$ lines respectively.
The agreement is significantly worse for the O lines, 24\% for the
O~{\small VIII}~Ly~$\alpha$ line and 23\% for the O~{\small VII}~For
line.  If only the CXO and XMM-Newton instruments are considered the
agreement improves.  The scaled normalizations agree to
within 4\% and 12\% for the  Ne~{\small IX}~Res and Ne~{\small
X}~Ly~$\alpha$ lines respectively. The discrepancy at  Ne~{\small
X}~Ly~$\alpha$ appears to be due mostly to the pn, with MOS and ACIS
in much better agreement ($\sim3\%$).  Again, the agreement is
significantly worse for the O lines, 16\% for the
O~{\small VIII}~Ly~$\alpha$ line and 22\% for the O~{\small VII}~For
line.  The discrepancy at  O~{\small VII}~For appears to be due mostly
to ACIS, with MOS and pn in much better agreement ($\sim3\%$).

   There are several possible reasons for the magnitude of the 
disagreement for these line normalizations. First, the effective area
for some of the instruments might be incorrect at some of the energies
examined in this analysis. This is particularly true for ACIS and the
XIS since both those instruments have time-variable contamination
layers which significantly affect the effective area at the O lines.
Second, the spectral redistribution function for some of the
instruments might be incorrect, leading to an incorrect computation of
the flux in a line.  Third, pileup may be distorting some of the
spectra leading to incorrect line normalizations.  We will be
exploring these possible explanations in the future.

\section{CONCLUSIONS}
\label{sect:conclusions} 

We have use the line-dominated spectrum of the SNR E0102 to test the
response models of the ACIS S3, MOS, pn, XIS, and XRT CCDs below
1.5~keV. We have fitted the spectra with the same model in which the
continuum and absorption components and the weak lines are held fixed
while allowing only the normalizations of four bright lines/line
complexes to vary.  We have compared the fitted line normalizations 
of the O~{\small VII}~For line, the O~{\small VIII}~Ly~$\alpha$ line, 
the Ne~{\small IX}~Res line, and Ne~{\small X}~Ly~$\alpha$ line
to examine the consistency of the effective area models for the
various instruments in the energy ranges around 570~eV, 654~eV,
915~eV, and 1022~eV.
We find that the instruments are in general 
agreement with 28 of the 32 scaled normalizations within $\pm10\%$
of the RGS determined values.
We find that the scaled line
normalizations agree to within 23\%, 24\%, 13\%, \& 19\% for
O~{\small VII}, O~{\small VIII}, Ne~{\small IX}, \& Ne~{\small X}
when all instruments are considered.  When only CXO and XMM-Newton are
considered, we find that the fitted line
normalizations agree to within 22\%, 16\%, 4\%, \& 12\% for
O~{\small VII}, O~{\small VIII}, Ne~{\small IX}, \& Ne~{\small X}.

%-------------
   \begin{figure}[tbh]
%   \vspace{-0.50in}
  \begin{center}
  \begin{tabular}{c}
  \includegraphics[width=3.5in,angle=270]{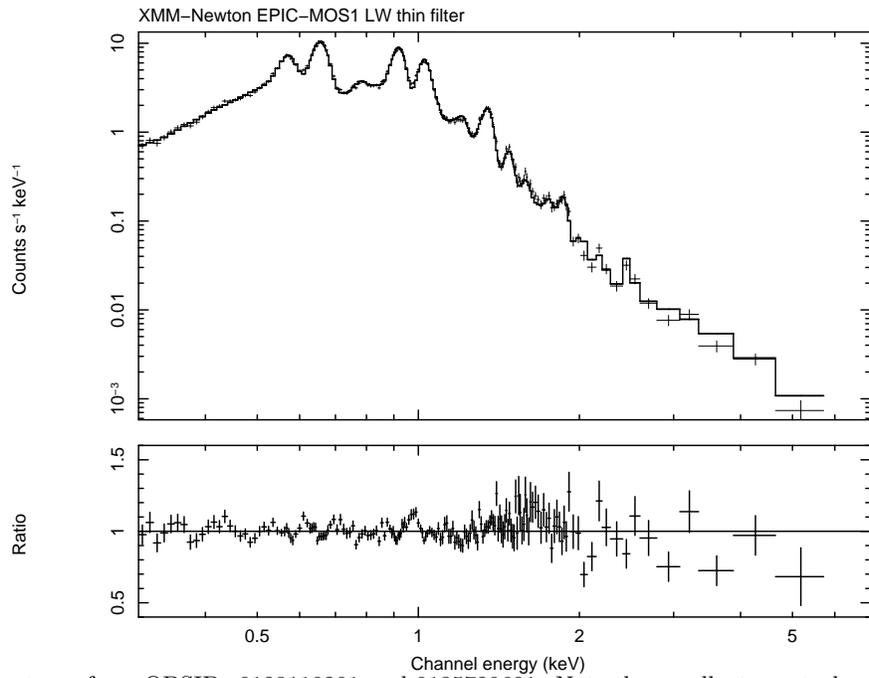}
  \end{tabular}
  \end{center}
  \vspace{-0.15in}
  \caption[MOS1] 
%>>>> use \label inside caption to get Fig. number with \ref{}
  { \label{fig:mos1_spec} MOS1 spectrum from OBSIDs 0123110201 and
0135720601. Note the excellent spectral resolution of the
MOS data. } 
  \end{figure} 
%-------------

%-------------
   \begin{figure}[bth]
%   \vspace{-0.50in}
  \begin{center}
  \begin{tabular}{c}
  \includegraphics[width=3.5in,angle=270]{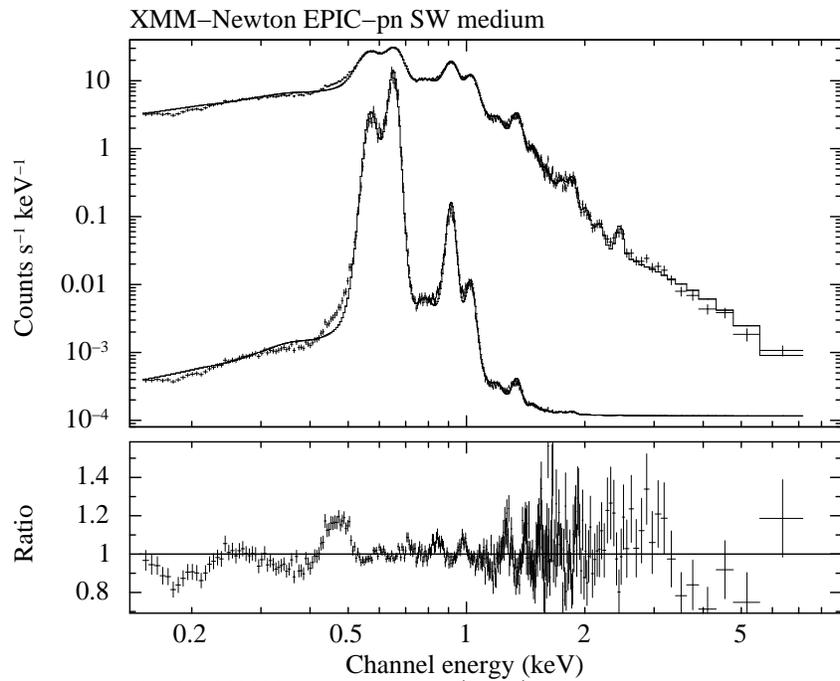}
  \end{tabular}
  \end{center}
  \vspace{-0.15in}
  \caption[pn] 
%>>>> use \label inside caption to get Fig. number with \ref{}
  { \label{fig:pn_spec} pn spectrum from OBSID0412980301 . The second (lower)
curve shows the same data but with a linear axis which has been
shifted downwards for clarity. Note the high count rate and the
pattern in the residuals which might indicate an issue with the
spectral redistribution function. } 
  \end{figure} 
%-------------

%-------------
   \begin{figure}[tbh]
%   \vspace{-0.50in}
  \begin{center}
  \begin{tabular}{c}
  \includegraphics[width=3.5in,angle=270]{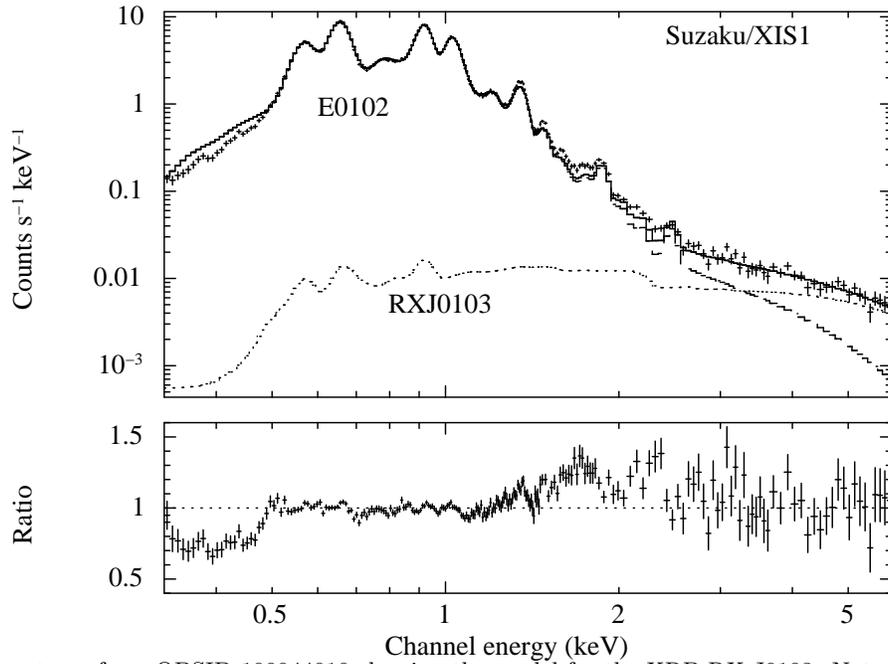}
  \end{tabular}
  \end{center}
  \vspace{-0.15in}
  \caption[pn] 
%>>>> use \label inside caption to get Fig. number with \ref{}
  { \label{fig:xis_spec} XIS1 spectrum from OBSID 100044010 showing
the model for the XRB RX~J0103. Note the residuals below 0.5~keV and
above 1.5~keV.} 
  \end{figure} 
%-------------

%-------------
   \begin{figure}[bth]
%   \vspace{-0.50in}
  \begin{center}
  \begin{tabular}{c}
  \includegraphics[width=3.5in,angle=270]{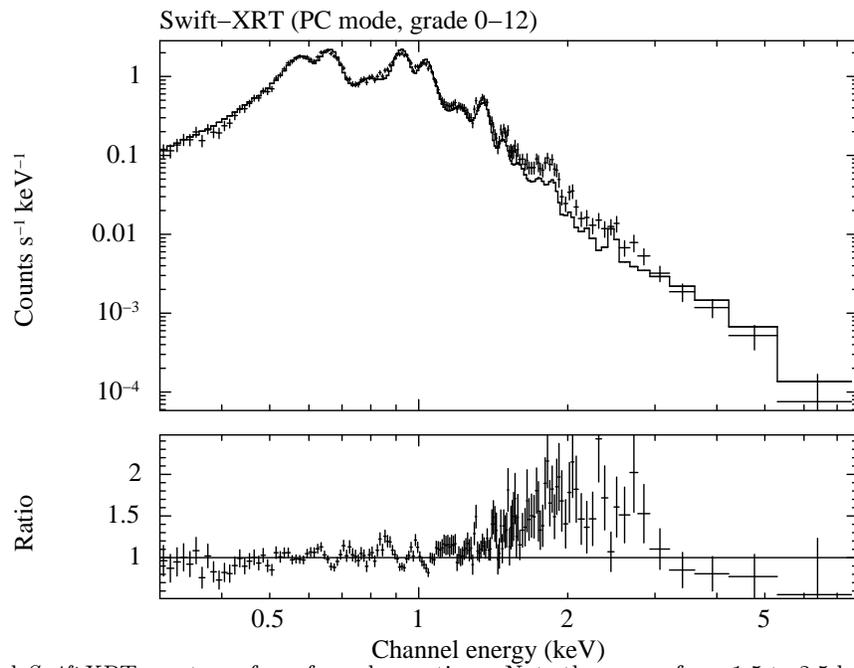}
  \end{tabular}
  \end{center}
  \vspace{-0.15in}
  \caption[pn] 
%>>>> use \label inside caption to get Fig. number with \ref{}
  { \label{fig:xrt_spec} Combined {\em Swift} XRT spectrum from four
observations.  Note the excess from 1.5 to 2.5 keV, this is a signature
of pileup. } 
  \end{figure} 
%-------------

%%%%%%%%%%%%%%%%%%%%%%%%%%%%%%%%%%%%%%%%%%%%%%%%%%%%%%%%%%%%%
%\newpage

\vspace*{-0.1in}
\acknowledgments     %>>>> equivalent to \section*{ACKNOWLEDGMENTS}       

%\vspace*{-0.1in}

This work was supported by NASA contract NAS8-03060.\\
We are grateful for the efforts of Marcus Kirsch in organizing and
leading the IACHEC which motivated the work which was presented in
this paper.

%We thank members of the CXC calibration group for enlightening discussions, 
%including Alexey Vikhlinin, Richard Edgar, Norbert Schulz, Dan Schwartz,  
%Herman Marshall, Glenn Allen, Dan Dewey and Katherine Flanagan.  
%We thank members of the ACIS instrument team for
%enlightening discussions including Gregory Prigozihn, Mark Bautz, Catherine 
%Grant, and Beverly LaMarr.
%We thank members of the XMM calibration teams for helping us with the XMM 
%data and data analysis including Frank Haberl, Andy Read, Steve
%Sembay, Michael Smith, Marcus Kirsch, Andy Pollock, and Martin Stuhlinger.
%For anyone we have forgotten, we apologize.

%We thank all of our colleagues on the CXO project who have
%contributed directly or indirectly to this work.  
%We thank
%Peter Ford,  Leisa Townsley, George Pavlov, Konstantin Getman, 
%and
%We thank all members of the ACIS
%instrument team, the CXC, the Chandra FOT, and MSFC Project Science 
%who have contributed to this
%effort.  

\clearpage
 
%%%%%%%%%%%%%%%%%%%%%%%%%%%%%%%%%%%%%%%%%%%%%%%%%%%%%%%%%%%%%
%%%%% References %%%%%

\bibliography{paper_ppp}   %>>>> bibliography data in report.bib
\bibliographystyle{spiebib}   %>>>> makes bibtex use spiebib.bst

\end{document}